\documentclass[preprint,12pt,3p,review]{elsarticle}
\usepackage[utf8]{inputenc} 
\usepackage[T1]{fontenc}    
\usepackage{hyperref}       
\usepackage{url}            
\usepackage{booktabs}       
\usepackage{amsfonts}       
\usepackage{nicefrac}       
\usepackage{microtype}      
\usepackage{lipsum}
\usepackage{multirow}
\usepackage{amsmath,amssymb,graphicx}
\usepackage{float}
\usepackage{relsize}
\usepackage{scalefnt}
\newcommand{\bigvarepsilon}{\mathlarger{\mathlarger{\varepsilon}}}
\newcommand{\bigsigma}{\mathlarger{\mathlarger{\sigma}}}
\usepackage{siunitx}
\usepackage{array}
\usepackage{lmodern}
\usepackage{amssymb}
\biboptions{sort&compress}
\usepackage{color}
\journal{Elsevier}
\begin{document}
\begin{frontmatter}
\title{A Computational Study On the Mechanical Properties of Pentahexoctite Single-layer: Combining DFT and Classical Molecular Dynamics Simulations}
\author[UFPI]{W. H. S. Brandão}
\author[UFPI]{A. L. Aguiar\corref{authorA}}
\cortext[authorA]{Corresponding authors}
\ead{acrisiolins@ufpi.edu.br}
\author[UnB]{L. A. Ribeiro Júnior}
\author[UNICAMP1,UNICAMP2]{D. S. Galvão}
\author[IFPI,UNICAMP1,UNICAMP2]{J. M. De Sousa\corref{authorA}}
\ead{josemoreiradesousa@ifpi.edu.br}

\affiliation[UFPI]{organization={Department pf Physics, Federal University of Piauí},
            addressline={Ininga}, 
            city={Teresina},
            postcode={64049-550}, 
            state={Piauí},
            country={Brazil}}
            
\affiliation[UnB]{organization={Institute of Physics, University of Brasília},
            city={Brasília},
            postcode={70910-900},
            country={Brazil}} 

\affiliation[UNICAMP1]{organization={Applied Physics Department, University of Campinas},
            addressline={Rua Sérgio Buarque de Holanda, 777 - Cidade Universitária}, 
            city={Campinas},
            postcode={13083-859}, 
            state={São Paulo},
            country={Brazil}}    

\affiliation[UNICAMP2]{organization={Center for Computing in Engineering and Sciences, University of Campinas},
            addressline={Rua Sérgio Buarque de Holanda, 777 - Cidade Universitária}, 
            city={Campinas},
            postcode={13083-859}, 
            state={São Paulo},
            country={Brazil}} 

\affiliation[IFPI]{organization={Federal Institute of Education, Science and Technology of Piau\'i -- IFPI},
            addressline={Primavera}, 
            city={São Raimundo Nonato},
            postcode={64770-000}, 
            state={Piauí},
            country={Brazil}}

\begin{abstract}
Studies aimed at designing new allotropic forms of carbon have received much attention. Recently, a new 2D graphene-like allotrope named Pentahexoctite was theoretically proposed. Pentahexoctite has a metallic signature, and its structure consists of continuous 5-6-8 rings of carbon atoms with $sp^{2}$ hybridization. Here, we carried out fully-atomistic computational simulations, combining reactive (ReaxFF) molecular dynamics (MD) and density functional theory (DFT) methods, to study the elastic properties and fracture patterns of Pentahexoctite monolayer. Results revealed a Young's Modulus of 0.74 TPa, smaller than the graphene one (about 1.0 TPa). The Pentahexoctite monolayer, when subjected to a critical strain, goes directly from elastic to completely fractured regimes. This process occurs with no plasticity stages between these two regimes. Importantly, graphene presents a similar fracture process. The elastic properties calculated with both DFT and MD are in good agreement.

\end{abstract}

\begin{keyword}
Reactive Molecular Dynamics \sep DFT \sep Mechanical Properties  \sep Carbon Allotrope \sep Pentahexoctite Sheet
\end{keyword}

\end{frontmatter}
\section{Introduction}

The accomplishments in producing the single-layer graphene (by mechanical exfoliation) resulted in the Nobel Prize in physics earned by Novoselov and Geim in 2004 \cite{novoselov2004}. Graphene is a one-carbon-atom-thick layer of graphite exhibiting unique properties, such as light-weight, good transparency, and high mechanical strength, and its advent gave rise to a new era in materials science \cite{sur2012graphene}. As a consequence of its success, several experimental \cite{toh2020synthesis,fan2021biphenylene} and theoretical \cite{wang2015phagraphene,wang2018popgraphene,zhang2015penta,li2017psi,jiang2017twin,doi:10.1002/pssb.201046583} studies have been carried out to propose new graphene-like allotropes. One objective of these studies is to propose new structures that can overcome some graphene limitations, such as high reactive to oxygen and null bandgap \cite{withers2010electron}.      

Based on that, a new 2D carbon allotrope graphene-like, named Pentahexoctite (PH), was recently proposed as a new metallic nanostructure composed of continuous 5-6-8 rings of carbon atoms with $sp^{2}$ hybridizations \cite{sharma2014pentahexoctite}. It was found that the Pentahexoctite presents a planar sheet with the absence of negative frequencies in the phonon spectra, which confirms its structural stability. They also reported that PH has mechanical strength comparable to graphene. In relation to the PH electronic properties, it exhibits metallic behavior with direction-dependent flat and dispersive bands at the Fermi level ensuring highly anisotropic transport properties. Its stability and electronic properties were well described in references \cite{sharma2014pentahexoctite,sui2017morphology}. However, its mechanical properties, elastic constants, and fracture patterns have not been fully investigated.

In this work, we carried out extensive fully-atomistic computational simulations using the reactive (ReaxFF) molecular dynamics (MD) and density functional theory methods to investigate the PH mechanical properties and fracture patterns. Our results show Young's Modulus of 247.93 GPa.nm (or 0.74 TPa). This value is comparable to the graphene one. The PH monolayer, when subjected to a critical strain, goes directly from elastic to completely fractured regimes. This trend is also exhibited by graphene. We also studied the impact of temperature on the PH fracture process. The results suggest that Young's Modulus varies between 0.82 to 0.50 TPa for temperatures ranging from 10 up to 1200 K, respectively. Importantly, the elastic properties calculated with both DFT and MD are in good agreement.

\begin{figure}[htb!]
    \centering
    \includegraphics[scale=0.4]{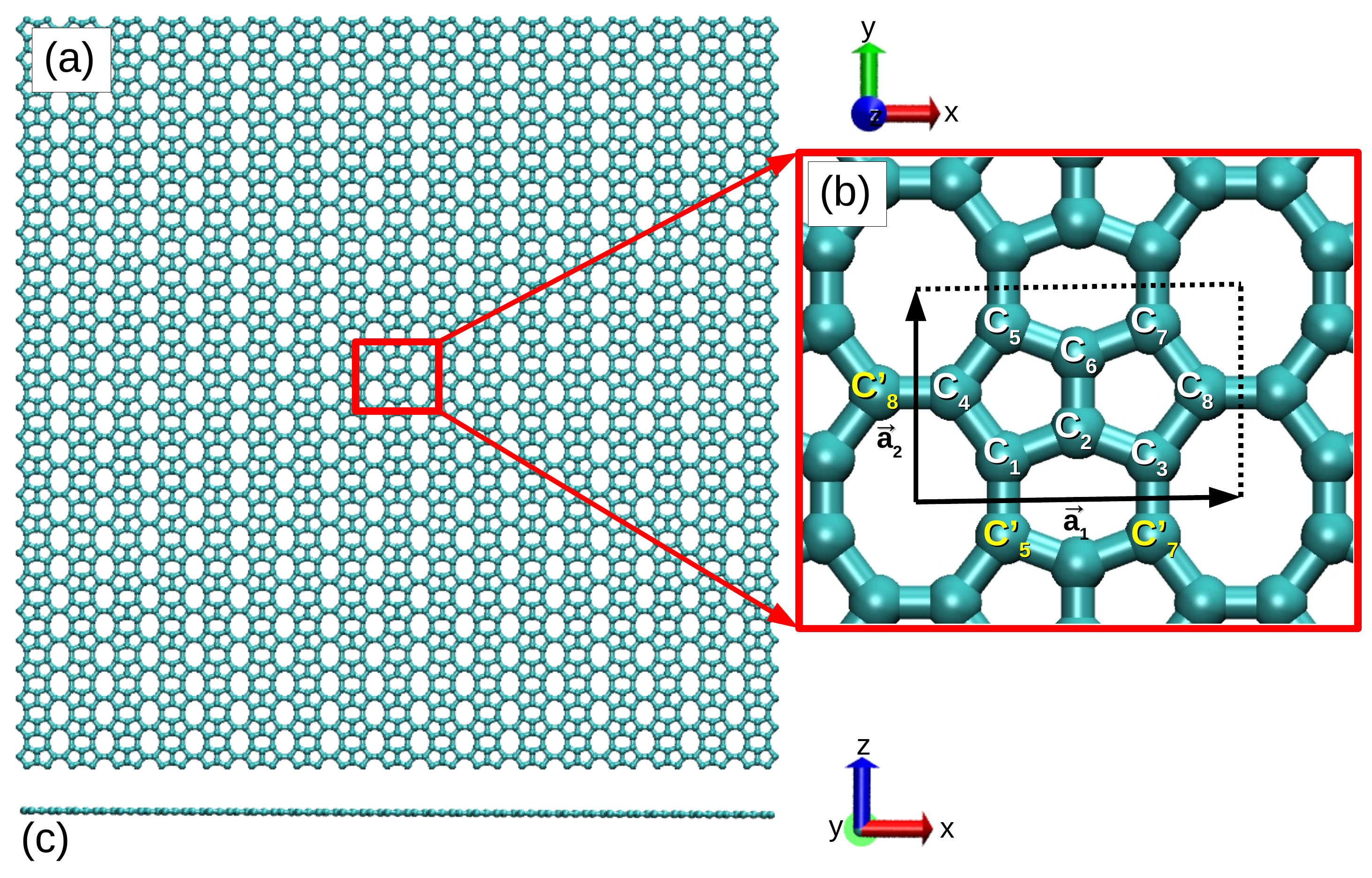}
    \caption{(a) Top view of the atomic model of Pentahexoctite monolayer (PH), (b) Zoomed view of its unit cell, and (c) its side view. The lattice vectors are $\mathbf{a_1}=5.85$ \r{A} $\hat{x}$ and $\mathbf{a_2}=3.78$ \r{A} $\hat{y}$. The bond distances are $C_1-C_2=1.43$ \r{A}, $C_1-C_4=1.49$ \r{A}, $C_2-C_6=1.37$ \r{A}, $C_1-C_5'=1.40$ \r{A}, and $C_4-C_8'=1.37$ \r{A} (where $C_i'$ is equivalent to $C_i$ atom).} 
    \label{fig:phog-supercell}
\end{figure}

\section{Computational Methodology}

\subsection{MD Method}

Here, we carried out extensive fully-atomistic MD/DFT simulations to investigate the PH mechanical properties. The classical reactive (ReaXFF) MD simulations were performed using the LAMMPS code \cite{plimpton1995fast}. ReaxFF is a modern force field designed to be a bridge between quantum and empirical chemical methods. It is an interatomic potential that allows the breaking ($C-C$ covalent bond) and formation of chemical bonds. It is appropriate for studying the elastic properties of nanostructures in the analysis of the mechanical failure. The ReaxFF parameter set used here is the one described in \cite{mueller2010development,van2001reaxff}. All MD snapshots presented here were rendered with the molecular visual dynamics (VMD) software \cite{humphrey1996vmd}.

PH has a 2D shape composed of 5-6-8 carbon rings, as illustrated in Figure 1. In this monolayer, the lattice vectors are $\mathbf{a_1}=5.85$ \r{A} $\hat{x}$ and $\mathbf{a_2}=3.78$ \r{A} $\hat{y}$. The bond distances are $C_1-C_2=1.43$ \r{A}, $C_1-C_4=1.49$ \r{A}, $C_2-C_6=1.37$ \r{A}, $C_1-C_5'=1.40$ \r{A}, and $C_4-C_8'=1.37$ \r{A}. In the MD simulations, it was subjected to uniaxial stresses along the x and y directions (see Figure 1). The temperature value ranges from 10 up to 1200 K. The PH structural model used in the MD simulations has a dimension 97 $\times$ 95 \r{A}, being composed of 3536 carbon atoms. Its thickness has approximately 3.34 \r{A} (i.e., one carbon atom thick). The used covalent interaction cutoff distance was $1.6$ \r{A}.  

In the MD simulations, Newton's equation of motion was integrated by velocity Verlet algorithm \cite{martys1999velocity}. The time step in all simulations is 0.05 fs. Before the stretching dynamics, the structure was thermalized using a Nose-Hoover thermostat during 10000 fs \cite{evans1985nose}. To eliminate any residual lattice stress before the stretching dynamics, an NPT simulation \cite{andersen1980molecular} was also carried out during 10000 fs. The elastic properties and fracture patterns at different temperatures (between 10-1200 K) were analyzed by stretching the PH structure along the x and y directions. We used a constant engineering tensile strain rate of $10^{-6}/fs$.

\begin{figure}[htb!]
    \centering
    \includegraphics[scale=0.5]{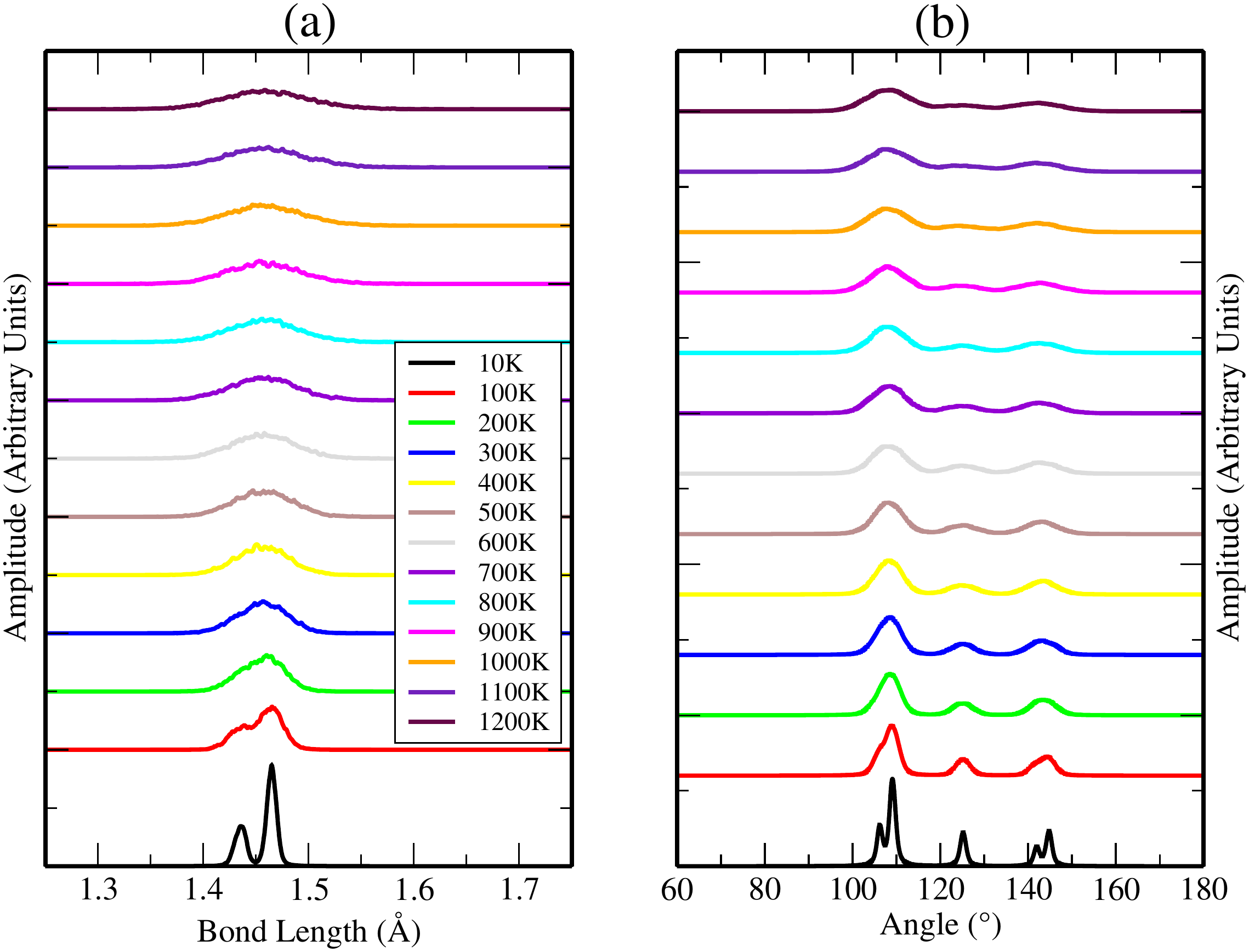}
    \caption{(a) PH bond and (b) angle distributions (in arbitrary units) as a function of temperature.} 
    \label{fig:bond-angle-dist}
\end{figure}

\subsection{DFT Method}

The DFT simulations were carried out using the SIESTA code \cite{soler2002}. The calculations were carried out within the scope of the generalized gradient approximation (GGA) with the exchange-correlation potential PBE \cite{pbe1996}. The Monkhorst-Pack \cite{mpack1976} grid of $8\times8\times1$ was considered in the self-consistent calculations. The double-$\zeta$ polarization (DZP) basis set was used. The standard value of 0.05 Ry for orbital cutoff \cite{pao2001} and the mesh cutoff of 400 Ry were also considered \cite{anglada2002}. We performed PH uniaxial stretching also through DFT calculations. A PH supercell was deformed considering a strain that ranged from 1 up to 30\%. This strain was applied by changing the lattice vectors corresponding to the stretching direction. The supercell was replicated 3$\times$5, which implies a lattice dimensions of approximately 17.55 \r{A} $\times$ 19.30 \r{A}, containing 120 atoms. The PH lattice was optimized using the tolerance for the force component of 0.01 ev/\r{A} at each strain step. Periodic boundary conditions were imposed, and the size of the box along the z-direction is equal to 40 \r{A}, which is large enough to avoid spurious effects due to the lattice images.


\begin{figure}[!htb]
    \centering
    \includegraphics[scale=0.45]{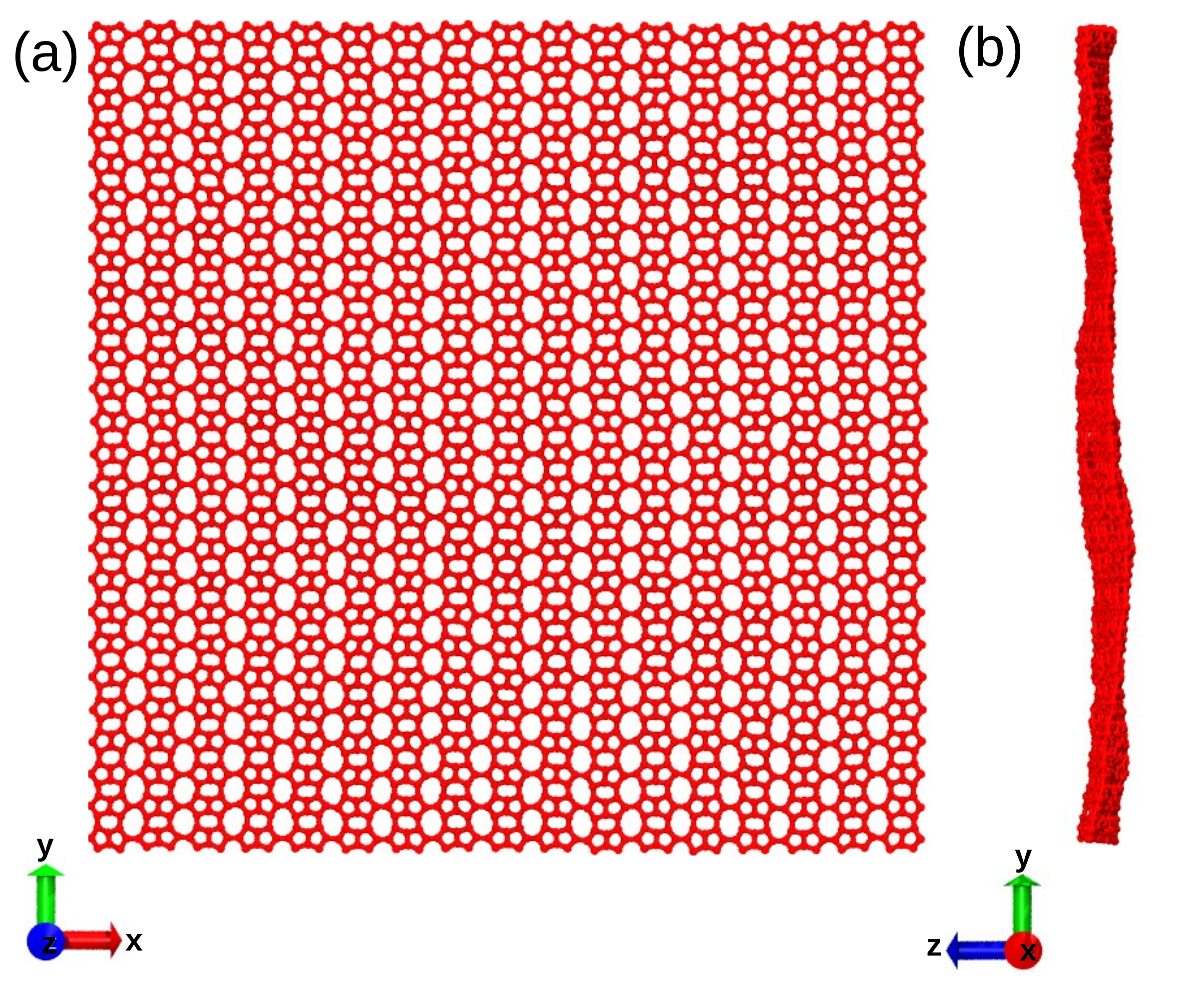}
    \caption{Representative MD snapshots for PH at 1200 K. (a) frontal and (b) lateral views.} 
    \label{fig:phog-1200K}
\end{figure}

\section{Results and Discussion}

We begin our discussion by presenting the PH bond (Figure  \ref{fig:bond-angle-dist} (a)) and angle (Figure  \ref{fig:bond-angle-dist} (b)) distributions after thermalization. As mentioned above, the monolayers were heated up in the range from 10 K to 1200 K. For low temperatures, we observed two peaks in the bond distribution pattern (Figure \ref{fig:bond-angle-dist}(a)) centered at 1.44 \r{A} and 1.46 \r{A}. This distribution pattern is consistent with the C--C bond length and they are within the range of the original values for the bond lengths shown in Figure 1 and reported in reference \cite{sharma2014pentahexoctite}. The increase in temperature leads to larger displacements of the carbon atoms that are imposed by thermal vibrations. In this way, the average bond length of the carbon atoms increases, and the two peaks that were formerly seen at 10 K are no longer distinguishable. This trend can be observed in the curves for temperatures ranging from 100 to 1200 K. However, even upon the structure heating, the new peak remains approximately between 1.4-1.5 \r{A} for all temperatures above 10 K. 

\begin{table}[htb!]
 \scalefont{0.5}
\centering
 \begin{tabular}{|c|c|c|c|c|}
 \multicolumn{5}{c}{x-direction} \\ \hline
 TEMP. (K) & $Y_{M}$(GPa.nm) & $US$(GPa.nm) & $\bigsigma_C$(GPa.nm) & $\bigvarepsilon_C$ \\ \hline
 DFT    & 237.26 $\pm$  3.51 & 26.03 $\pm$ 0.71 & 26.03 $\pm$ 0.71 & 0.15 \\ \hline
 10	    &  268.08 $\pm$   1.77	 &  13.68 $\pm$   0.01	 &  13.68 $\pm$  0.04	&   0.085	\\ \hline
 100	&  260.36 $\pm$   0.68	 &  13.18 $\pm$   0.04	 &  13.07 $\pm$  0.04	&   0.085	\\ \hline
 200	&  249.41 $\pm$   0.90	 &  12.19 $\pm$   0.07	 &  12.00 $\pm$  0.05	&   0.084	\\ \hline
 300	&  239.11 $\pm$   1.22	 &  12.03 $\pm$   0.08	 &  11.81 $\pm$  0.04	&   0.071	\\ \hline
 400	&  233.33 $\pm$   1.01	 &  11.79 $\pm$   0.07	 &  11.75 $\pm$  0.05	&   0.064	\\ \hline
 500	&  222.44 $\pm$   1.77	 &  12.04 $\pm$   0.07	 &  12.04 $\pm$  0.04	&   0.063	\\ \hline
 600	&  212.78 $\pm$   2.05	 &  11.65 $\pm$   0.09	 &  11.51 $\pm$  0.03	&   0.063	\\ \hline
 700	&  206.98 $\pm$   1.93	 &  11.86 $\pm$   0.03	 &  11.68 $\pm$  0.04	&   0.067	\\ \hline
 800	&  200.49 $\pm$   2.03	 &  11.28 $\pm$   0.04	 &  11.28 $\pm$  0.04	&   0.062	\\ \hline
 900	&  191.07 $\pm$   2.29	 &  11.10 $\pm$   0.04	 &  10.90 $\pm$  0.05	&   0.065	\\ \hline
 1000	&  188.15 $\pm$   2.14	 &  10.23 $\pm$   0.03	 &  10.22 $\pm$  0.07	&   0.059	\\ \hline
 1100	&  173.81 $\pm$   2.47	 &  10.63 $\pm$   0.09	 &  10.02 $\pm$  0.08	&   0.065	\\ \hline
 1200	&  167.56 $\pm$   2.38	 &  10.50 $\pm$   0.09	 &  10.48 $\pm$  0.08	&   0.065   \\ \hline
 \multicolumn{5}{c}{y-direction} \\ \hline
TEMP. (K) & $Y_{mod}$(GPa.nm) & UTS(GPa.nm) & $\bigsigma_C$(GPa.nm) & $\bigvarepsilon_C$ \\ \hline
DFT & 261.88 $\pm$ 3.62 & 33.26 $\pm$ 0.99 & 33.26 $\pm$ 0.99 & 0.19 \\ \hline
  10	&  261.25 $\pm$   7.22	 &  17.87 $\pm$   0.06	 &  17.87 $\pm$  0.06	&   0.128	\\ \hline
 100	&  258.25 $\pm$   5.92	 &  17.07 $\pm$   0.08	 &  17.07 $\pm$  0.05	&   0.121	\\ \hline
 200	&  258.98 $\pm$   4.25	 &  15.59 $\pm$   0.17	 &  15.56 $\pm$  0.03	&   0.114	\\ \hline
 300	&  256.75 $\pm$   3.02	 &  14.83 $\pm$   0.15	 &  14.78 $\pm$  0.02	&   0.124	\\ \hline
 400	&  245.27 $\pm$   2.98	 &  13.81 $\pm$   0.24	 &  13.71 $\pm$  0.01	&   0.119	\\ \hline
 500	&  242.09 $\pm$   2.12	 &  13.01 $\pm$   0.08	 &  12.87 $\pm$  0.04	&   0.114	\\ \hline
 600	&  234.84 $\pm$   1.93	 &  12.76 $\pm$   0.28	 &  12.72 $\pm$  0.05	&   0.093	\\ \hline
 700	&  227.95 $\pm$   1.81	 &  12.68 $\pm$   0.52	 &  12.57 $\pm$  0.03	&   0.091	\\ \hline
 800	&  220.61 $\pm$   1.71	 &  12.30 $\pm$   0.04	 &  12.21 $\pm$  0.02	&   0.081	\\ \hline
 900	&  214.58 $\pm$   1.96	 &  12.12 $\pm$   0.35	 &  11.97 $\pm$  0.05	&   0.079	\\ \hline
 1000	&  209.38 $\pm$   1.55	 &  11.82 $\pm$   0.06	 &  11.67 $\pm$  0.04	&   0.073	\\ \hline
 1100	&  196.76 $\pm$   2.26	 &  11.50 $\pm$   0.47	 &  11.15 $\pm$  0.07	&   0.073	\\ \hline
 1200	&  196.05 $\pm$   2.09	 &  11.00 $\pm$   0.60	 &  10.82 $\pm$  0.04	&   0.070	\\ \hline
  \end{tabular}
  \caption{PH elastic constants for both x and y-directions.}
  \label{tab:elastic-values}
  \end{table}

Importantly, we did not observe any bond break or new structural arrangements for temperatures up to 1200 K, as illustrated in Figure \ref{fig:phog-1200K} (a). This fact suggests that the PH is structurally stable in this temperature range. However, we can observe the loss of the initial planar character (see Figure \ref{fig:phog-1200K} (b)) The angle distribution for low temperatures reveals five peaks around 106$^\circ$, 109$^\circ$, 125$^\circ$, 142$^\circ$, and 145$^\circ$. But the heating spread the peaks, and only three of them can be distinguished around 109$^\circ$, 125$^\circ$ and 144$^\circ$. These results further support that PH is also dynamically stable, corroborating previous results reported in the literature \cite{sharma2014pentahexoctite} through phonon spectra calculations.


\begin{figure}[!htb]
    \centering
    \includegraphics[scale=0.5]{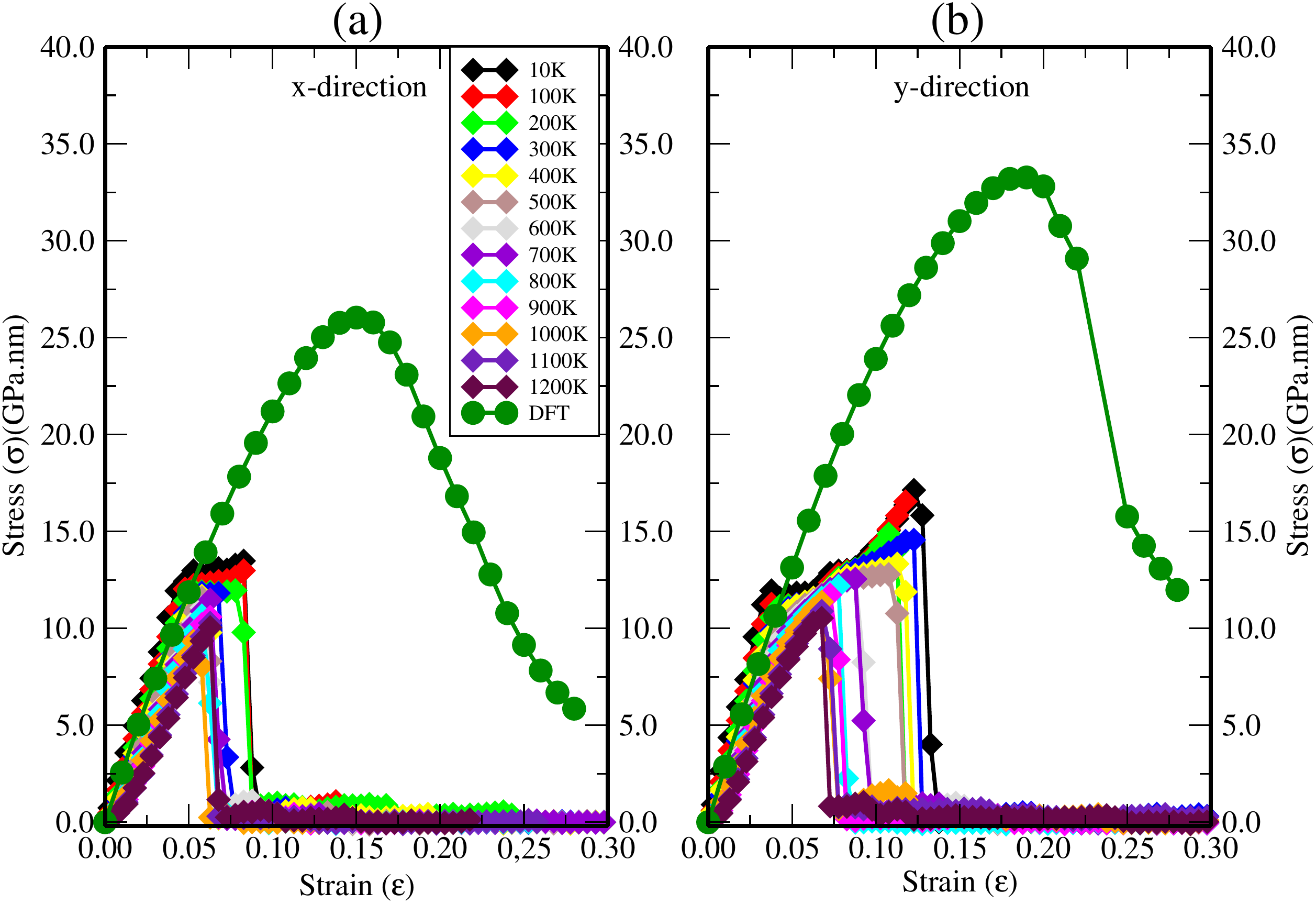}
    \caption{PH stress-strain curves under uniaxial tensile loading along the (a) x-direction, and (b) y-direction. The green curves indicate the DFT results.}
    \label{fig:ss-PHOG}
\end{figure}

We discuss now the PH mechanical properties calculated through MD and DFT simulations. Figure \ref{fig:ss-PHOG} shows the stress-strain curves for temperatures varying from 10 up to 1200 K obtained from the MD simulations, considering uniaxial tensile loading applied along the \ref{fig:ss-PHOG}(a) x and \ref{fig:ss-PHOG}(b) y-directions. The green curves in Figure \ref{fig:ss-PHOG} show the corresponding DFT results. The first noticeable feature in this figure is that PH has a significant degree of anisotropy regarding the fracture strain (i.e., the critical strain value, $\bigvarepsilon_C$). PH is more resilient to tensile loading along the y-direction than along the x-one. Another result that we can observe from Figure \ref{fig:ss-PHOG} is the difference in the Ultimate Stress ($US$) presented for the stress-strain curves for the different methodologies. The DFT curves (green) have a higher $US$ when contrasted to the MD ones for both directions. We attribute the observed differences to the distinct methodologies used in the DFT and MD simulations. For example, DFT results are for zero temperatures. Furthermore, the stretching process used in the MD simulations is dynamic, while in the DFT approach, it is static and based on a simulation protocol that involves stretching-relaxing-stretching the sheet to a predetermined strain value. Therefore, we believe that successive PH relaxations to energy minimum states, under a given strain value, have contributed to increasing its $US$.  


\begin{figure}[htb!]
    \centering
    \includegraphics[scale=0.4]{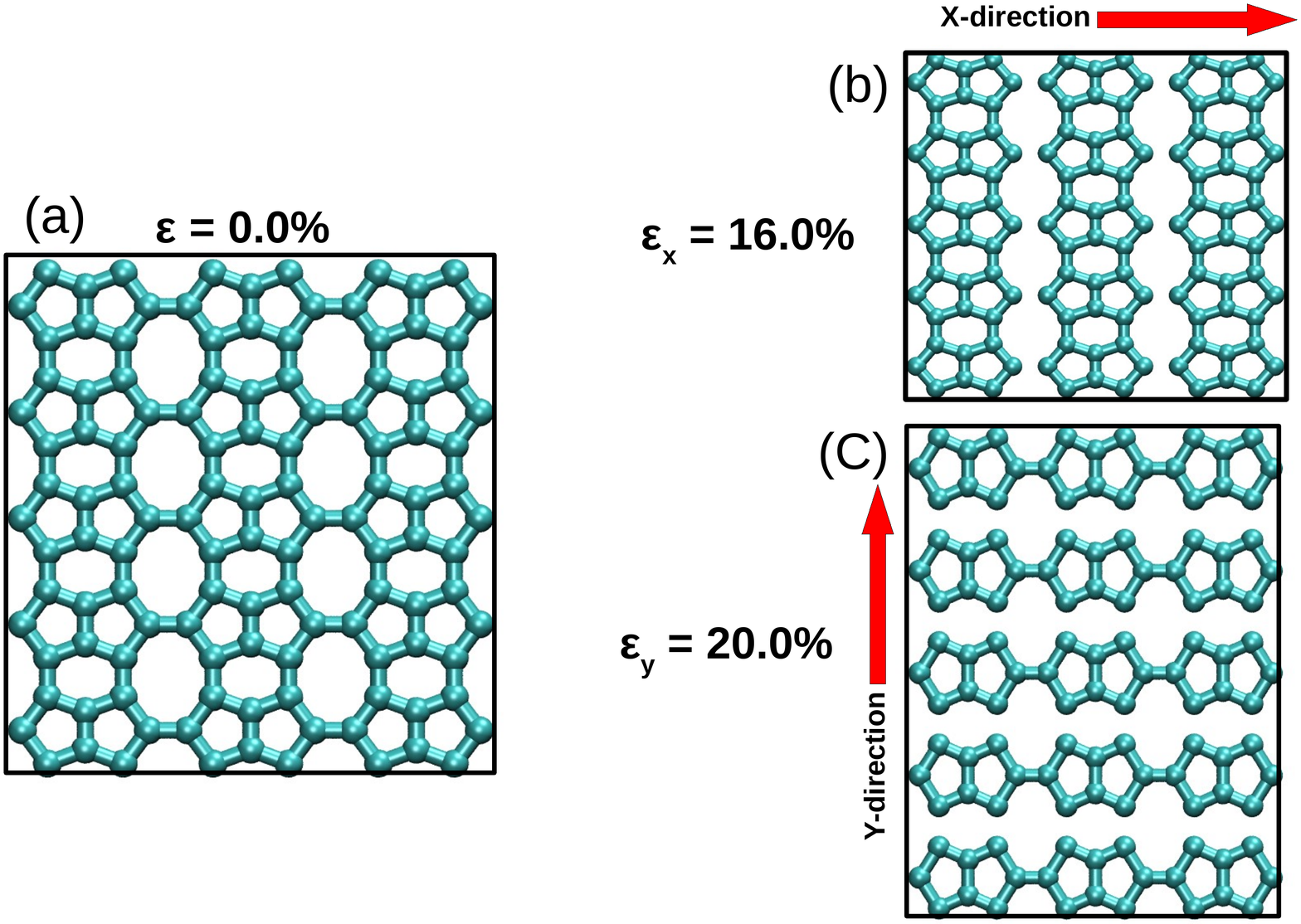}
    \caption{PH lattice arrangements under uniaxial strain at (a) 0\%, (b) 16\% strain along the x-direction, and (c) 20\% strain along the y-direction.}
    \label{fig:membrane-strain}
\end{figure}

We summarize the results for the elastic constants in Table \ref{tab:elastic-values}. In our calculations, 0.05 strain (linear regime) was used to estimate Young's Modulus ($Y_{M}$) values. The ultimate stress ($US$) is defined as the corresponding tensile stress for a critical strain ($\varepsilon_{critical}$). In all MD simulations, the critical strain is characterized by the structural failure (fracture). We calculated the virial stress $\sigma_C$ along the stretching directions according to the approach described in reference \cite{brandao2021atomistic}. $Y_{M}$ changes significantly with increasing temperature, about 100 GPa along the x-direction, and by approximately 70 GPa along the y-one. $Y_{M}$ values for 300, 600, 900, and 1200 K are close to results obtained with another graphene-like allotrope that has a lattice structure similar to PH, i.e., the popgraphene monolayer \cite{junior2020temperature}). Like $Y_{M}$, the $US$, $\bigsigma_C$, and $\bigvarepsilon_C$ decrease smoothly with increasing temperature. These last quantities are slightly larger for the y-direction than for the x-one. This can be attributed to the different orientations, along the x and y-directions, of the PH $C--C$ bonds. We can see from Figure \ref{fig:phog-supercell}(b) that there is one bond parallel to the x-direction ($C_4-C'_8$) and another four bonds quasi-parallel to it ($C_1-C_2$, $C_2-C_3$, $C_5-C_6$ and $C_6-C_7$). On the other hand, there are two bonds parallel to the y-direction ($C_1-C'_5$ and $C_3-C'_7$).

\begin{figure}
    \centering
    \includegraphics[scale=0.6]{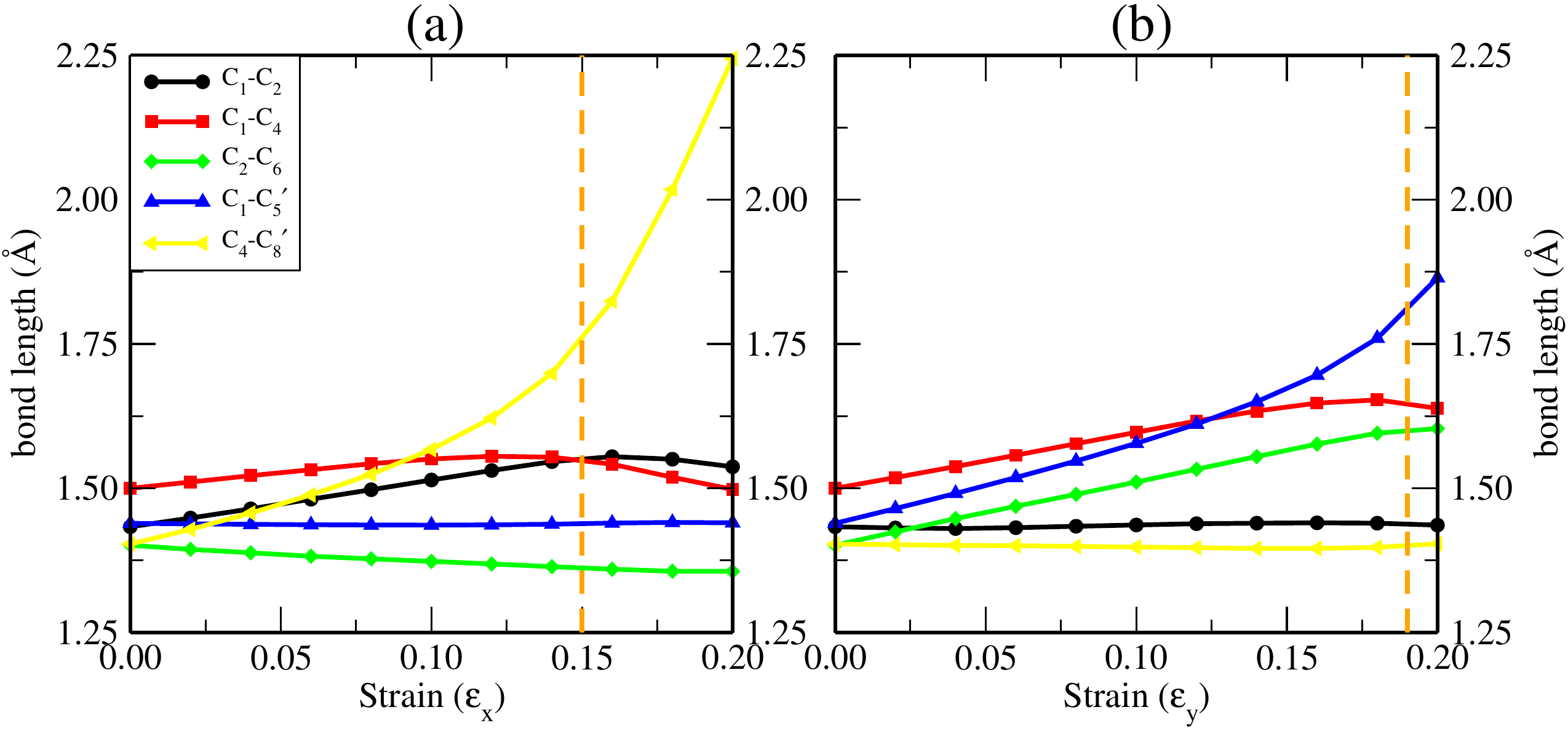}
    \caption{Bond length evolution as a function of the applied strain along (a) x-direction (a) and (b) y-direction. Each curve refers to a bond type presented in the PH unit-cell, as shown in Figure \ref{fig:phog-supercell}(b).}
    \label{fig:bond-ev}
\end{figure}

The PH fracture process from the DFT simulations is shown in Figure \ref{fig:membrane-strain}. Figure \ref{fig:membrane-strain}(a) illustrates the monolayer at null stress. In Figure \ref{fig:membrane-strain}(b), We can note that when the uniaxial strain is along the x-direction, the breaking bonds involve the ones that are parallel to the stretching direction (i.e., bond types $C_4$-$C'_8$ ). On the other hand, in the case of the y-direction stretching (see Figure \ref{fig:membrane-strain}(c)), the fracture starts from the bonds parallel to this direction (i.e., bond types $C_2-C_6$).


\begin{figure}[!htb]
\centering
\includegraphics[scale=0.4]{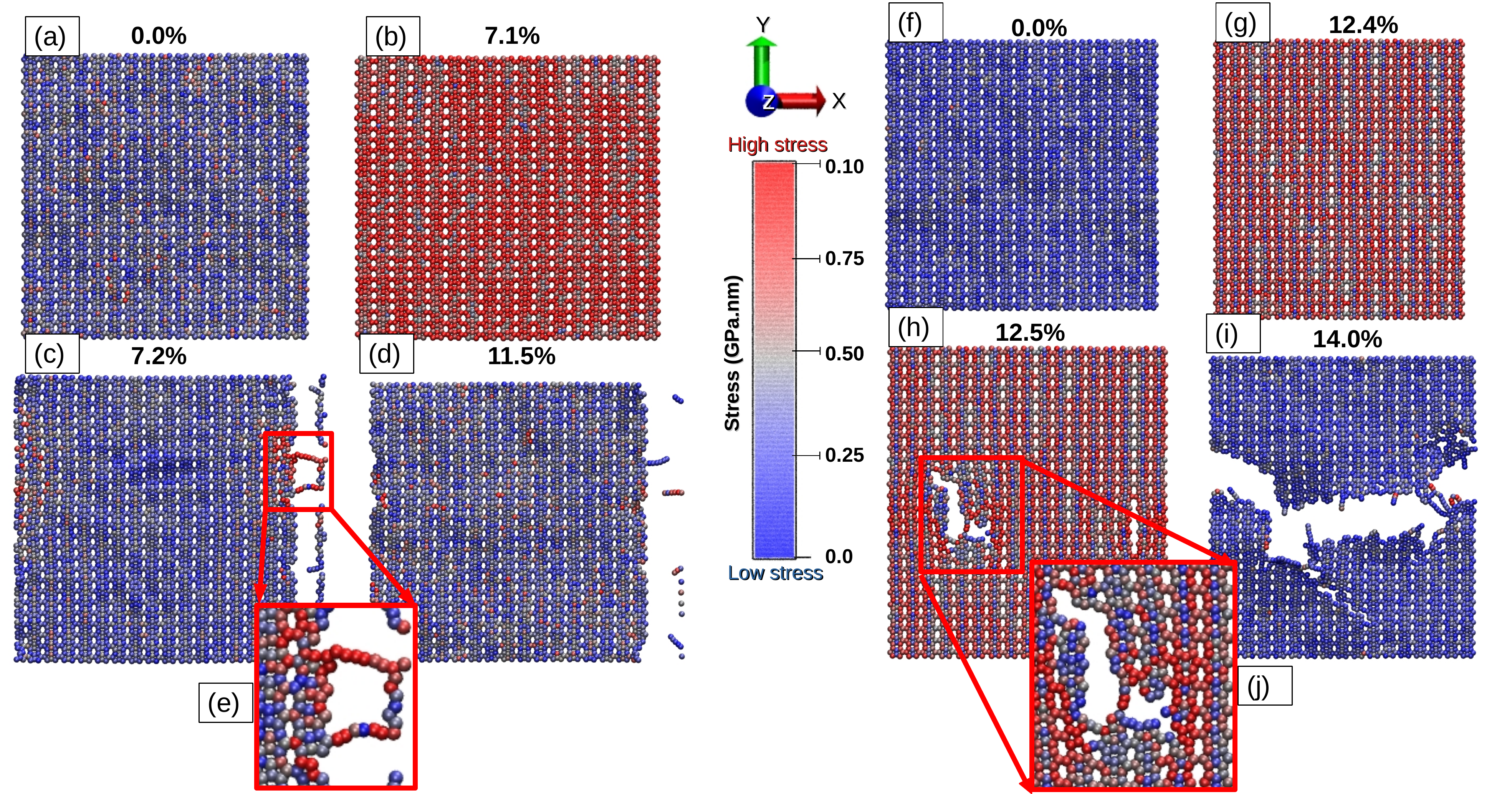}
\caption{Representative MD snapshots for the uniaxial tensile loading simulations for the strain applied along the (a-e) x-direction and (f-j) y-direction. The color scheme indicated the von Mises stress per-atom values according to the value bar. Red and blue colors indicate high- and low-stress accumulation, respectively.}
\label{fig:300K-x}
\end{figure} 

The PH bond lengths evolution as a function of the applied uniaxial tensile strain is shown in Figure \ref{fig:bond-ev}. The bond $C_4$-$C_8'$ (parallel to the x-direction strain) experiences the largest variation in length and are related to the fracture at 15\% strain (orange vertical traced line in \ref{fig:bond-ev}(a)). However, the $C_1$-$C_5'$ bond (perpendicular to the stretching direction) has practically no variation in length. It is interesting to note that the $C_2$-$C_6$ bond, perpendicular to the axis of the applied strain, has a slight decrease in its length. For the y-direction case (Figure \ref{fig:bond-ev}(b)), the bonds $C_1$-$C_4$, $C_2$-$C_6$ and $C_1$-$C_5'$ experience considerable variations in length, the largest being the bond $C_1$-$C_5'$ (parallel to the y-direction). The bonds $C_1$-$C_2$ and $C_4$-$C_8'$ remain practically constant throughout the tensile process. Bond fracture is seen at 19\% strain (dashed vertical orange line in Figure \ref{fig:bond-ev}(b)).


Finally, we analyzed the PH fracture pattern from the MD simulations. Figure \ref{fig:300K-x} shows representative MD snapshots for the fracture process at 300K. Figures \ref{fig:300K-x}9(a-e) and \ref{fig:300K-x}9(f-j) indicate the cases for the stretching along the x and y-directions, respectively. The color scheme in these figures represents the spatial distribution of the von Mises (VM) stress per-atom values \cite{mises_1913}. Red and blue colors indicate the high- and low-stress accumulation, respectively. Importantly the VM values provide useful information on the fracture process since they reveal the fracture point or region. Details about the VM calculations can be found in the LAMMPS manuals \cite{lammps_manual} and in the reference \cite{brandao2021atomistic}.  

From Figure \ref{fig:300K-x}, we can see that PH when subjected to a critical strain, goes directly from elastic to completely fractured regimes. This process occurs with no plasticity stages between these two regimes. Importantly, graphene presents a similar fracture process \cite{felix2020mechanical}. In the case of the tensile loading along the x-direction, the first bonds to break are the ones parallel to the direction of the applied strain (see Figure \ref{fig:300K-x}(e)). After that, the lattice structural collapse occurs abruptly nearby one of the edges at 6.85 \% strain, as shown in Figure \ref{fig:300K-x}(d). Since the lattice has periodic boundary conditions, one can see the fragmentation and the formation of linear atomic chains (LACs) on the right edge, whereas on the left one, we can see the appearance of vacancies and dangling bonds. In the case of the tensile loading along the y-direction (see Figures \ref{fig:300K-x}(f-j)), we can see that the first stages of the lattice induce the formation of a crack (i. e, a fractured region in (see Figures \ref{fig:300K-x}(h))) instead of a single bond breaking. The crack propagation is fast, and the lattice becomes separated into two parts at around 11.5\% of strain. In this case, we can also see the LAC formation. 


\section{Conclusions}

In summary, we have investigated the mechanical properties and fracture patterns of a pentahexoctite monolayer (PH) subjected to different temperature regimes combining DFT and reactive (ReaxFF) MD simulations. Our results reveal that the PH is thermally stable up to 1200 K. The elastic constants obtained with DFT are approximately twice larger than those obtained with MD. The exception is Young's modulus which has a good agreement between both approaches. We obtained the following average values for Young's Modulus, Ultimate Strenght, and critical strain in MD simulations: 740.09 TPa (247.93 GPa.nm), 40.09 TPa (13.43 GPa.nm), 9.75\% at 300 K. These values gradually decrease with increasing temperature. An anisotropy for the PH mechanical behavior was observed. For the y-direction, we obtained higher values for the elastic constants when compared to the x-direction. PH when subjected to a critical strain, goes directly from elastic to completely fractured regimes. This process occurs with no plasticity stages between these two regimes. Importantly, graphene presents a similar fracture process. The crack propagation is fast and the lattice becomes separated into two parts with the presence of linear atomic chains (LACs).  

\section{Acknowledgements}

This work was supported in part by the Brazilian Agencies CAPES, CNPq, FAPDF, and FAPESP. L.A.R.J and J.M.S acknowledges CENAPAD-SP (Centro Nacional de Alto Desenpenho em São Paulo - Universidade Estadual de Campinas - UNICAMP) for computational support process (proj842 and proj634). A.L.A. acknowledges CNPq (Process No. $427175/20160$) for financial support. W.H.S.B., A.L.A. and J.M.S  thank the Laboratório de Simulação Computacional Cajuína (LSCC) at Universidade Federal do Piau\'i for computational support. L.A.R.J acknowledges the financial support from a Brazilian Research Council FAP-DF grants $00193-00000857/2021-14$, $00193-00000853/2021-28$, and $00193-00000811/2021-97$ and CNPq grant $302236/2018-0$, respectively. L.A.R.J. gratefully acknowledges the support from ABIN grant 08/2019. L.A.R.J. acknowledge N\'ucleo de Computaç\~ao de Alto Desempenho (NACAD) for providing the computational facilities through the Lobo Carneiro supercomputer. L.A.R.J. thanks Fundaç\~ao de Apoio \`a Pesquisa (FUNAPE), Edital 02/2022 - Formul\'ario de Inscriç\~ao N.4, for the financial support. A.F.F thanks the Brazilian Agency CNPq for Grant No. 303284/2021-8 and São Paulo Research Foundation (FAPESP) for Grant No. \#2020/02044-9. This research also used the computing resources and assistance of the John David Rogers Computing Center (CCJDR) in the Institute of Physics “Gleb Wataghin”, University of Campinas.

\newpage

 \bibliographystyle{elsarticle-num} 
 \bibliography{cas-refs}
\end{document}